%% file: rosepo.tex
\documentclass[sigconf]{acmart}

\AtBeginDocument{%
  }

\setcopyright{acmlicensed}
\copyrightyear{2018}
\acmYear{2018}
\acmDOI{XXXXXXX.XXXXXXX}

\acmConference[Conference acronym 'XX]{Make sure to enter the correct
  conference title from your rights confirmation emai}{June 03--05,
  2018}{Woodstock, NY}
\acmISBN{978-1-4503-XXXX-X/18/06}

\usepackage{multirow}
\usepackage{booktabs}
\usepackage{color}
\definecolor{lightgray}{RGB}{215,215,215}
\usepackage{colortbl}
\usepackage{xcolor}
\usepackage{subcaption}
\usepackage{footnote}

\usepackage{enumitem}       

\newcommand{\ie}{\emph{i.e., }}
\newcommand{\eg}{\emph{e.g., }}

\newcommand{\cf}{\emph{cf. }}

\begin{document}

\title{RosePO: Aligning LLM-based Recommenders with Human Values}

\author{Jiayi Liao}
\email{ljy0ustc@mail.ustc.edu.cn}
\orcid{0009-0006-7998-8462}
\affiliation{
  \institution{University of Science and Technology of China}
  \city{Hefei}
  \country{China}
}

\author{Xiangnan He}
\authornote{Corresponding authors.}
\email{xiangnanhe@gmail.com}
\orcid{0000-0001-8472-7992}
\affiliation{%
  \institution{University of Science and Technology of China}
  \city{Hefei}
  \country{China}
}

\author{Ruobing Xie}
\authornotemark[1]
\email{xrbsnowing@163.com}
\orcid{0000-0003-3170-5647}
\affiliation{%
  \institution{Machine Learning Platform Department, Tencent}
  \city{Beijing}
  \country{China}
}

\author{Jiancan Wu}
\affiliation{%
  \institution{University of Science and Technology of China}
  \city{Hefei}
  \country{China}
}

\author{Yancheng Yuan}
\affiliation{%
  \institution{The Hong Kong Polytechnic University}
  \city{Hong Kong}
  \country{China}
}

\author{Xingwu Sun}
\affiliation{%
  \institution{Machine Learning Platform Department, Tencent}
  \city{Beijing}
  \country{China}
}

\author{Zhanhui Kang}
\affiliation{%
  \institution{Machine Learning Platform Department, Tencent}
  \city{Shenzhen}
  \country{China}
}

\author{Xiang Wang}
\affiliation{%
  \institution{University of Science and Technology of China}
  \city{Hefei}
  \country{China}
}

\renewcommand{\shortauthors}{Trovato et al.}

\input{chapters/abstract}

\ccsdesc[500]{Information systems~Recommender systems}

\keywords{Large Language Models, Sequential Recommendation, Preference Optimization}


\maketitle

\input{chapters/intro}
\input{chapters/related}
\input{chapters/method}
\input{chapters/exp}
\input{chapters/conclusion}


\bibliographystyle{ACM-Reference-Format}
\bibliography{rosepo}

\clearpage
\input{chapters/appendix}

\end{document}

%% file: chapters/abstract.tex
\begin{abstract}
Recently, there has been a growing interest in leveraging Large Language Models (LLMs) for recommendation systems, which usually adapt a pre-trained LLM to the recommendation scenario through supervised fine-tuning (SFT).
However, both the pre-training and SFT stages fail to explicitly model the comparative relationships of a user's preferences on different items.
To construct a ``helpful and harmless'' LLM-based recommender, we propose a general framework --- \textbf{R}ec\textbf{o}mmendation with \textbf{s}moothing p\textbf{e}rsonalized \textbf{P}reference \textbf{O}ptimization (\textbf{RosePO}), which better aligns with customized human values during the post-training stage.
Specifically, in addition to the input and chosen response that naturally align with SFT data, we design a rejected sampling strategy tailored for enhancing helpfulness, along with two strategies aimed at mitigating biases to promote harmlessness.
To ensure robustness against uncertain labels present in automatically constructed preference data, we introduce a personalized smoothing factor predicted by a preference oracle into the optimization objective.
Evaluation on three real-world datasets demonstrates the effectiveness of our method, showcasing not only improved recommendation performance but also mitigation of semantic hallucination and popularity bias.
\end{abstract}

%% file: chapters/intro.tex
\section{Introduction}
Sequential recommendation aims to predict users' subsequent interest in items, by analyzing their historical interactions. 
Inspired by the extensive knowledge and reasoning abilities demonstrated by Large Language Models (LLMs) \cite{gpt-4, llama-3, claude, qwen2} across various domains \cite{chatlaw, bloomberggpt, med-llm-survey}, LLM for recommendation (LLM4Rec) has gradually gained significant attention \cite{tallrec, p5, chatrec, instructrec}.
To adapt a general-purpose LLM for sequential recommendation, a common paradigm \cite{llara, bigrec, transrec} involves two key steps:
(1) Transform a sequence of historical user-item interactions into a text-like input prompt, with the subsequent interacted item as the supervised response, and then convert this input-response pair into a next-item-prediction instruction;
(2) Apply supervised fine-tuning (SFT) on the converted data, leveraging autoregressive modeling on the response tokens \cite{gpt-3, instructgpt}.
After SFT, LLM is capable of following instructions to perform recommendation tasks like next-item prediction.

\begin{figure}[t]
\centering
\includegraphics[width=0.46\textwidth]{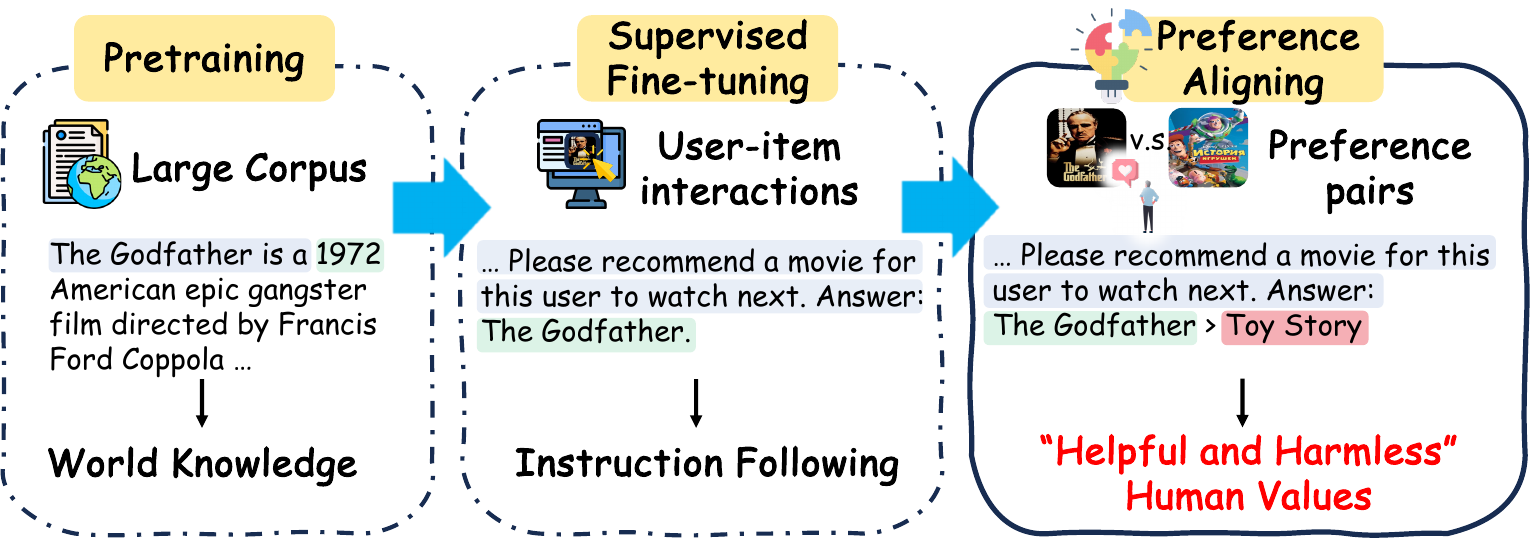}
\vspace{-10pt}
\caption{Training of an LLM-based recommender.}
\vspace{-12pt}
\label{fig:teaser}
\end{figure}

However, solely using the subsequent item as the chosen response (\ie positive label) falls short in developing an ``HH'' (\ie \textit{helpful} and \textit{harmless} \cite{claude, 3h}) recommender system.
By helpful, we mean that a recommender model should accurately distinguish the next item of interest from irrelevant items, thus enhancing the user experience by providing relevant suggestions \cite{mf, ncf}.
Nonetheless, the point-wise SFT paradigm only fits the chosen responses, neglecting the finer distinctions between item preferences.
By harmless, we mean that a recommendation service should be free from biases, whether inherent in the LLM or present in the data, thus promoting fairness and inclusivity \cite{bias_rs, fairness_rs}.
Yet, the SFT paradigm primarily focuses on fitting observed historical behaviors without explicitly addressing or mitigating potential biases.

To this end, we explore pairwise preference learning in recommendation \cite{bpr} --- differentiating between chosen and rejected responses.
This idea naturally aligns with direct preference optimization (DPO) \cite{dpo}, which holds promise for achieving helpfulness and harmlessness \cite{claude, agent_alignment}.
This necessity raises a critical research question: ``How could we leverage DPO to develop an HH recommender system?''
In Natural Language Processing (NLP), DPO typically follows two steps:
(1) Construct offline preference data, which consists of input prompt, chosen response, and rejected response; (2) Optimize the comparison between the chosen and rejected responses given an input.
By adapting this two-step pipeline into LLM4Rec, we propose a new framework --- \textbf{R}ec\textbf{o}mmendation with \textbf{s}moothing p\textbf{e}rsonalized Preference Optimization (\textbf{RosePO}), which constructs preference pairs towards HH and performs personalized smoothing optimization for robustness.

\begin{figure*}[t]
\centering
\includegraphics[width=1.0\textwidth]{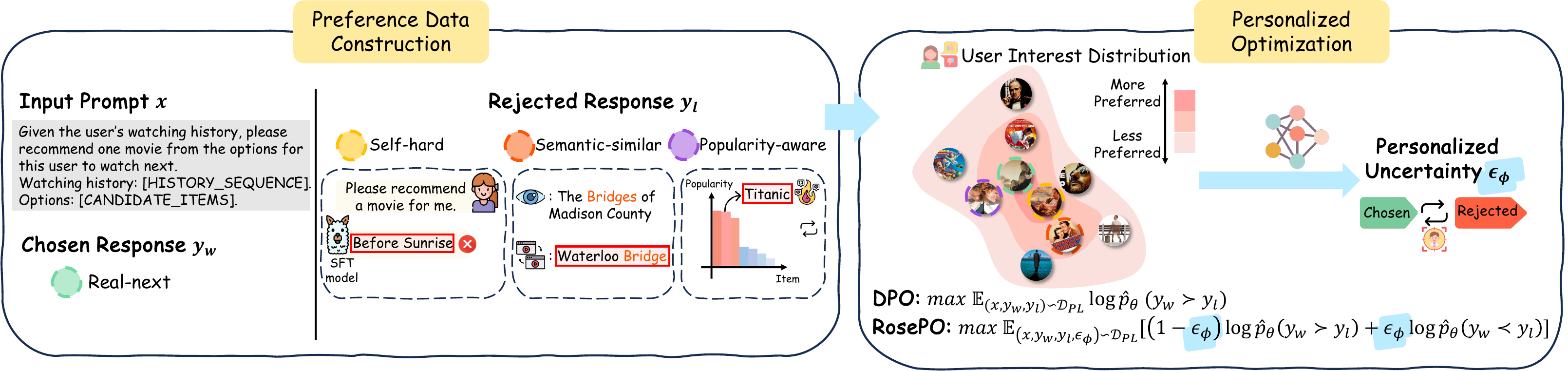}
\vspace{-20pt}
\caption{The framework of RosePO comprises two main components: preference data construction and personalized optimization. (1) The rejected items are sampled based on specific preferences for HH. (2) We estimate the uncertainty of preference for each data point guided by a preference oracle, and inject this personalized uncertainty as a smoothing factor into the optimization objective.}
\vspace{-10pt}
\label{fig:method}
\end{figure*}

\vspace{5pt}
\noindent
\textbf{Construction of Preference Pairs}.
Naturally, the input and chosen response remain consistent with prompts describing the sequence of historical items and the target item of interest during the SFT stage.
As the input and chosen responses already exist, the key to designing the preference pairs lies in selecting the rejected responses.
Instead of randomly selecting uninteracted items as the rejected responses, we draw inspiration from hard negative sampling \cite{ns_theory} and design three strategies: one to improve helpfulness and two to enhance harmlessness.
\begin{itemize}[leftmargin=*]
    \item \textbf{Addressing self-hard cases.}
    To accurately differentiate the chosen item from challenging candidates, we create self-hard rejected responses by sampling from the SFT model's incorrect predictions (\eg the movie ``Before Sunrise'' in Figure \ref{fig:method}).
    This allows the model to learn from its mistakes \cite{self-correct}, thereby enhancing the prediction accuracy of LLM-recommenders.
    \item \textbf{Mitigating semantic hallucination.}
    The SFT model might focus too much on semantic similarities, which can lead to misleading recommendations \cite{homogeneity_issue}.
    To tackle this, we propose selecting items (\eg the movie ``Waterloo Bridge'' in Figure \ref{fig:method}) with high semantic similarity to the historical sequence (\eg this user has watched the movie ``The Bridges of Madison Country'') as rejected samples.
    This helps the model avoid making assumptions based solely on surface-level similarities.
    \item \textbf{Reducing popularity bias.}
    Fine-tuning on interaction data might introduce popularity bias, potentially reducing fairness for both users and items \cite{popularity_bias}.
    To address this, we suggest selecting rejected samples (\eg the movie ``Titanic'' in Figure \ref{fig:method}) based on item popularity.
    This helps counteract the popularity bias and promotes a more balanced recommendation service.
\end{itemize}

\begin{figure}[t]
\centering
\includegraphics[width=0.47\textwidth]{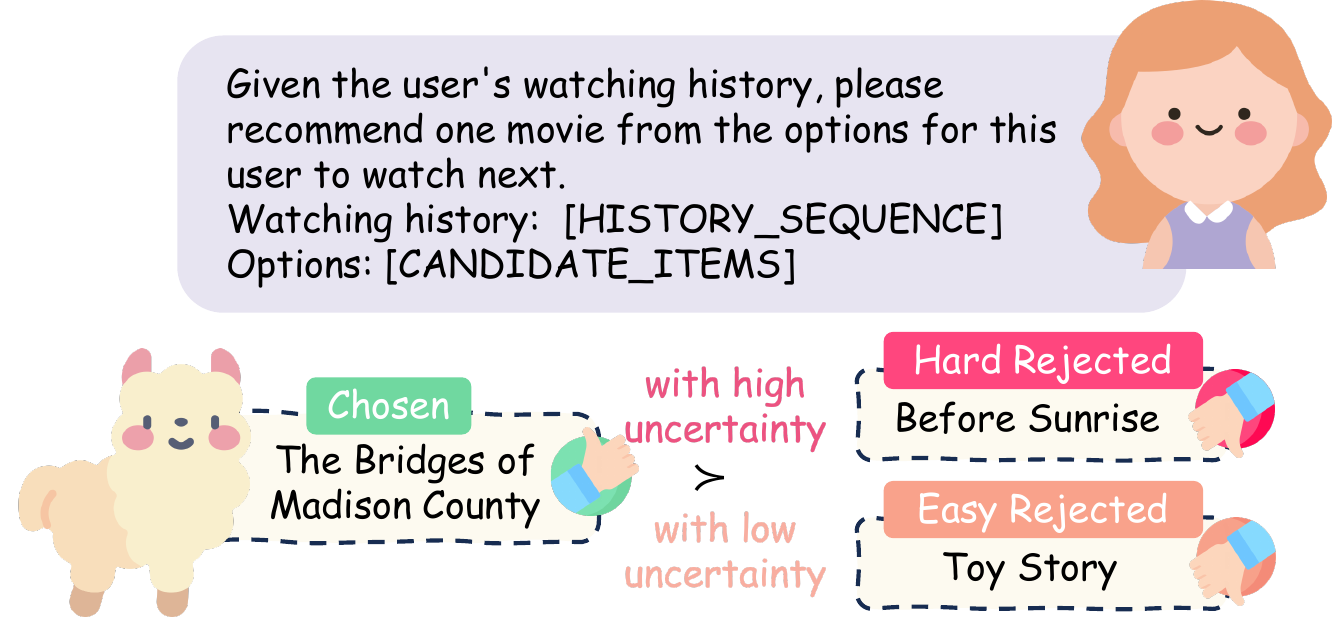}
\vspace{-5pt}
\caption{Since this user loves romantic movies, the preference for her subsequent interaction with ``The Bridges of Madison County'' over ``Before Sunrise'' exhibits greater uncertainty than over ``Toy Story''.}
\vspace{-5pt}
\label{fig:dpo}
\end{figure}

\vspace{5pt}
\noindent
\textbf{Personalized Preference Optimization}.
Unlike in the NLP field, where preference data is directly annotated by humans, preference learning in recommendation faces a significant challenge: the observed interactions between users and items are sparse \cite{mf}.
This raises the risk of selecting false negatives in the rejected samples, with the likelihood of false negatives increasing as the samples become harder.
Consequently, it is uncertain that users will exhibit a higher interest in the chosen items compared with the rejected ones, and the preference gap between them varies (\cf Figure~\ref{fig:dpo}).
To address this, we introduce personalized uncertainty into the optimization objective, which assesses the likelihood that a user prefers the chosen response over the rejected one.
This uncertainty is derived from the predictions of the probabilities for different users to interact with both chosen and rejected items, which are approximated by a lightweight auxiliary recommendation model.
By incorporating this new objective, we can achieve more robust preference optimization in recommendation scenarios, particularly in the presence of noisy labels \cite{cdpo, rdpo}, thereby allowing the model to generalize across a broader range of data quality.

In summary, our contributions are threefold:
(1) We propose a general framework capable of handling various preferences in developing a helpful and harmless recommendation service.
(2) By introducing a personalized uncertainty term into DPO loss, we enhance the model's robustness for noisy preferences in automatically-constructed labels.
(3) Empirical evaluations on three datasets and various scenarios demonstrate the effectiveness of our approach.

%% file: chapters/related.tex
\section{Related Work}
In this section, we present a literature review of LLMs for Recommendation and preference alignment in LLMs, from which our research draws inspiration.

\subsection{LLMs for Recommendation}
In this era of information explosion, to alleviate users from being overloaded by massive products and content, sequential recommendation aims to mine users' interests based on their historical behavior sequences, filtering information and selecting next items for them to interact with.
Amazing at the emergent abilities of LLMs, including the memorization of world knowledge and reasoning capabilities, recent works \cite{chatrec, instructrec, p5, recformer} explore their potential in the field of recommendation (LLM4Rec), the core of which is transforming recommendation tasks into language tasks.

Existed studies leverage LLMs for recommendation from three main aspects: 
(1) LLMs as recommender: LLMs can make recommendation decisions based on users' past engagement \cite{tallrec, llara};
(2) LLMs as enhancer: LLMs can enhance traditional recommendation models by processing and generating text features of users and items \cite{morec, unisrec};
(3) LLMs as simulator: LLMs can empower agents to simulate users in recommendation scenarios \cite{agent4rec, agentcf, recmind}. 
Scrutinizing the first paradigm, LLM can be converted into a recommender through either prompting (\eg zero-shot \cite{zero-shot-ranker}, in-context learning \cite{demorec}, or chain-of-thought \cite{drdt}) or tuning (\eg parameter-efficient fine-tuning \cite{lora} or full fine-tuning).
Going a step further, our work discusses how to introduce preference in tuning LLM as a sequential recommender.
Different from \cite{dmpo, softmax_dpo}, which increases the number of negatives in DPO within LLM4Rec, our focus is on selecting appropriate negatives based on preferences and optimizing the pairwise loss function in a personalized smoothing manner.

\subsection{Preference Alignment in LLMs}
Benefiting from the scaling law \cite{scaling_laws}, LLMs \cite{gpt-4, llama-3} exhibit excellent performance in solving tasks across different domains \cite{chatlaw, bloomberggpt, med-llm-survey}.
The development of LLMs usually comprises two stages, pre-training and post-training \cite{llama-3}.
Pre-training with the next-token-prediction task on a huge amount of corpora can acquire knowledge, while post-training can improve LLMs' capabilities in following instructions and preferences.
In the post-training stage, RLHF \cite{rlhf, instructgpt, claude} aligns LLM with human values by reinforcement learning with a reward model representing human feedback.

Since the online RLHF is hard to optimize, offline algorithms are explored to improve stability and simplicity.
Take the well-known DPO \cite{dpo} as an example, it extracts the optimal policy in a closed form and directly optimizes LLM with preference data.
Various optimization objectives are also designed: several works try to sample preference pairs from the optimal policy \cite{rpo, rso}; some develop reference-free objectives \cite{cpo, simpo}; others aim at more robust tuning \cite{ipo, cdpo, rdpo, beta-dpo}.
Different from optimizing a general preference in the NLP field, we optimize for personalized preferences with noisy labels in the recommendation scenario.

%% file: chapters/method.tex
\section{Method: RosePO \includegraphics[width=0.025\textwidth]{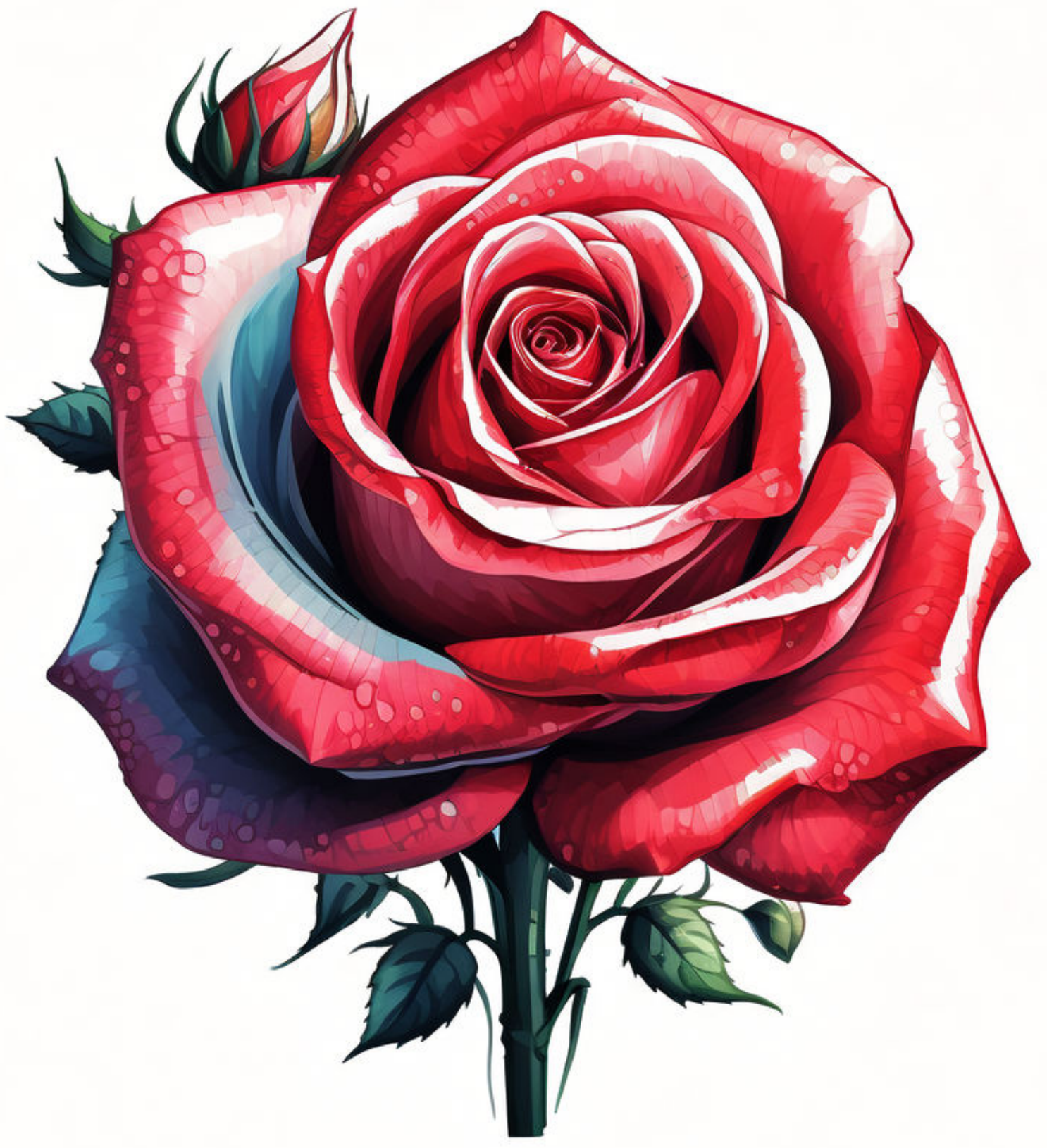}}
\subsection{Background}
\vspace{5pt}
\noindent{\textbf{Task Formulation.}}
A user $u$ chronically interacted with items $i_1, i_2, \cdots i_n$.
Sequential recommendation aims to predict the next item $i_{n+1}$ that the user will be interested in based on the sequence of historical items.

\vspace{5pt}
\noindent{\textbf{Supervised Fine-Tuning of LLM4Rec.}}
To adapt a general LLM for recommendation tasks, a typical SFT procedure for domain adaptation involves two steps:
\begin{enumerate}[leftmargin=*]
    \item Present the user-item interaction data into the text-like format as finetuning dataset $\mathcal{D}_\text{SFT}=\left\{(x,y)\right\}$.
    Here, input $x$ requires the LLM to select an item from a given candidate set based on the user's historical interaction sequence, while response $y$ corresponds to the target item that the user will interact with.
    \item Fine-tune the LLM parameterized by $\theta$ with autoregressive modeling.
    The objective is to maximize the likelihood of the chosen response $y$ given the input prompt $x$:
    \begin{equation}\max_{\theta} \mathbb{E}_{(x,y)\thicksim\mathcal{D}_\text{SFT}} \pi_{\theta}(y \mid x). \label{eq:sft}
    \end{equation}
\end{enumerate}

\vspace{5pt}
\noindent{\textbf{Direct Preference Optimization.}}
During the stage of preference alignment in LLMs, Direct Preference Optimization (DPO) \cite{dpo} extracts a closed-form optimal policy of RLHF \cite{rlhf} and model the reward function as:
\begin{equation}\begin{aligned}r(x,y)=\beta\log\frac{\pi_\theta(y\mid x)}{\pi_\text{ref}(y\mid x)}+\beta\log Z(x), \label{eq:reward} \end{aligned}\end{equation}
where the partial function is $\begin{aligned}Z(x) = \sum_y\pi_{\text{ref}}(y | x)\exp\left(\frac{1}{\beta}r(x,y)\right)\end{aligned}$ and $\beta$ represents the deviation of policy $\pi_\theta$ from the reference model $\pi_{\mathrm{ref}}$.
For the preference learning dataset $\mathcal{D}_\text{PL}=\left\{(x,y_w,y_l)\right\}$, with $x$, $y_w$, and $y_l$ denoting the input prompt, chosen response, and rejected response, respectively, preference can be formulated by Bradley-Terry model \cite{bradley_terry} as:
\begin{equation}
\begin{aligned}
\hat{p}(y_w\succ y_l\mid x) &= \sigma(r(x,y_w)-r(x,y_l)) \\
&= \sigma\left(\beta\log\frac{\pi_\theta(y_w\mid x)}{\pi_{\text{ref}}(y_w\mid x)} - \beta\log\frac{\pi_\theta(y_l\mid x)}{\pi_{\text{ref}}(y_l\mid x)}\right),
\label{eq:BT}
\end{aligned}\end{equation}
where $\sigma$ refers to the sigmoid function.
Then, DPO optimizes the preference model as :
\begin{equation}\max_{\theta} \mathbb{E}_{(x,y_w,y_l)\thicksim\mathcal{D}_\text{PL}}\log \hat{p}(y_w\succ y_l\mid x).\label{eq:bc_loss}\end{equation}

When applying DPO to the SFT model in the recommendation domain, we face two key challenges:
(1) How to identify preference pairs, $y_w$ and $y_l$, that can accurately reflect users' personalized preferences, while ensuring both helpfulness and harmlessness?
(2) How to tackle the various levels of uncertainty in preference labels (\ie $y_w\succ y_l \mid x$) and optimize the DPO loss accordingly?
We will address these two challenges in Sections \ref{sec:data_construction} and \ref{sec:preference_optimization}, respectively.

\subsection{Construction of Preference Pairs} \label{sec:data_construction}
Regarding the first challenge, we focus on three types of preference pairs (illustrated on the left side of Figure \ref{fig:method}): one for helpfulness and two for harmlessness.

\vspace{5pt}
\noindent
\textbf{Data format}.
Preference learning for LLM-based recommenders requires transforming user-item interaction data into a preference format.
Formally, we need to elaborate the following three key elements in the preference data $\mathcal{D}_\text{PL}=\left\{(x,y_w,y_l)\right\}$:
\begin{enumerate}[leftmargin=*]
    \item \textbf{Input prompt $x$.} This includes a task description of sequential recommendation, a user's interaction history with items, and candidate items.
    \item \textbf{Chosen response $y_w$.} 
    We treat the user's selected item $i_{n+1}$ following the interaction sequence as the chosen response, regardless of the sampling strategy used for the rejected response.
    \item \textbf{Rejected response $y_l$.} 
    This is a sampled negative item, which is the central focus of this paper.
\end{enumerate}

Consider Figure \ref{fig:method} as an example.
Wherein, the chosen response $y_w$ is ``The Bridges of Madison County'', while the rejected response $y_l$ can be any item except the chosen one.
Note that both the chosen response $y_w$ and the rejected one $y_l$ are involved among the candidate items in the input prompt $x$, from which the model is expected to make a choice.

\vspace{5pt}
\noindent
\textbf{Rejected samples for helpfulness}.
Current SFT practices in recommendation \cite{bigrec, transrec, e4srec, llara} rarely incorporate negative examples in the explicit responses, often relying on random candidates in the input prompt.
This implicit modeling overlooks pairwise learning at the level of response, which might negatively impact performance.
This motivates us to apply pairwise preference learning between the chosen (positive) and rejected (negative) responses.
Furthermore, inspired by studies on dynamic negative sampling (DNS) in recommendation \cite{dns, ns_theory}, we perform uniform sampling among candidate items with higher predictive probabilities by the SFT model than the selected item $i_{n+1}$.
As illustrated in Figure \ref{fig:method}, we select ``Before Sunrise''  falsely recommended by the SFT model to create the self-hard rejected responses.
Just like correcting mistakes in one's own error collection, minimizing the probability of generating these items can help the model rectify its own deficiencies.

\vspace{5pt}
\noindent
\textbf{Rejected samples for harmlessness}.
In addition to focusing on helpfulness, other preferences can also be incorporated into our framework to enhance harmlessness.
Here we take two representative preferences that users usually care about as examples.

\begin{itemize}[leftmargin=*]
\item \textbf{Mitigating semantic hallucination. }
LLM-based recommenders often rely on semantic correlations in historical sequences to infer user preferences \cite{llara}.
However, excessive reliance on these correlations can lead to spurious associations.
To address this issue, we create preference pairs designed to mitigate semantic hallucination.
Formally, let $[\text{emb}_1]$, $[\text{emb}_2]$, $\cdots$, $[\text{emb}_n]$ represent the textual embeddings of historical items $i_1$, $i_2$, $\cdots$, $i_n$, respectively.
We define the semantic similarity between item $i_k$ and the historical sequence as
$\sum_{t=1}^n\frac{\text{CosSim}([\text{emb}_t], [\text{emb}_k])}{n}$, where $\text{CosSim}(\cdot)$ refers to cosine similarity.
We select the item $i^*$ from the whole item set with the highest semantic similarity to the user's historical sequence as the rejected response.
By applying preference optimization over such preference pairs, distinguishing the target response from these rejected items is expected to avoid spurious semantic correlations.
\item \textbf{Mitigating popularity bias. }
Exploration and diversity are essential in recommendation systems, as they enhance users' long-term satisfaction \cite{fairness_and_diversity_in_rs}.
However, as demonstrated in Figure~\ref{fig:bias}, pure SFT often leads to popularity bias\cite{popularity_bias} in LLM-based recommender.
For example, the popularity bias of the SFT model is as high as 24.58\% for Goodreads dataset.
To address this issue, we construct preference pairs by sampling popular items as rejected responses based on item popularity distribution.
By optimizing on such preference pairs, the ability to differentiate the target response from popular items can contribute to the alleviation of popularity bias.
\end{itemize}

\subsection{Personalized Preference Optimization} \label{sec:preference_optimization}

Having established the preference pairs based on our three strategies, we now tackle the second obstacle: How can we handle the varying levels of uncertainty in preference labels and optimize the DPO loss accordingly?
Specifically, the rejected responses --- particularly those sampled using the hard strategy --- may actually represent false negatives \cite{false_neg}.
Directly applying DPO on such preference pairs could lead to detrimental outcomes. 
To tackle this, we incorporate a measure of uncertainty into the DPO loss function.
This uncertainty reflects the likelihood of noisy labels that a user's actual preference may differ from the predefined labels (\ie $y_w\succ y_l \mid x$), especially when false negatives are present (\ie $y_w\prec y_l \mid x$).
By adjusting the DPO loss to account for the uncertainty, we can mitigate the risk of misinterpreting false negatives.

Formally, Equation \eqref{eq:bc_loss} regresses $\hat{p}(y_w\succ y_l \mid x)$ to the target $p(y_w\succ y_l \mid x)=1$ by minimizing the cross entropy, which implies that $r(x,y_w)-r(x,y_l) \to+\infty$ in Equation \eqref{eq:BT}.
However, unlike human-annotated preference data, the preference data we automatically construct in the recommendation context may not fully adhere to the assumption $p(y_w\succ y_l \mid x)=1$, indicating the presence of noisy labels.
Inspired by the conservative DPO loss \cite{cdpo}, we introduce a smoothing factor $\epsilon$, to quantify the uncertainty of label flipping.
That is, we modify the target to $p(y_w\succ y_l \mid x)=1-\epsilon$ with an $\epsilon$ term representing the uncertainty regarding the preference comparison.
Then, the conservative optimization objective can be expressed as: 
\begin{equation}
\min_\theta \mathbb{E}_{(x,y_w,y_l)\thicksim\mathcal{D}_\text{PL}} \mathcal{L}_{\text{CE}}(\hat{p}_\theta(y_w\succ y_l\mid x), p_\theta(y_w\succ y_l\mid x)), 
\label{eq:RosePO}
\end{equation}
where $\mathcal{L}_{\text{CE}}(\cdot,\cdot)$ is the cross entropy loss and the target is updated to $p(y_w\succ y_l \mid x)=1-\epsilon$.

Furthermore, we argue that using a static hyperparameter $\epsilon$ is inadequate for dynamically adjusting based on the difficulty of each case.
For instance, ``Before Sunrise'' is more perplexing as a rejected response than ``Toy Story'' for a user with a history of enjoying romantic movies, while ``Before Sunrise'' is less confusing as a rejected response than ``Toy Story'' for a child who loves animated films.
Therefore, we propose predicting personalized label uncertainty, denoted as $\epsilon_\phi$, for each data point (as illustrated on the right side of Figure \ref{fig:method}).
This process utilizes a preference oracle, $\phi$, which is capable of predicting the probability of user $u$ interacting with an item $i_k$ next, based on the user's historical sequence. 

Specifically, we represent the probability predictions of $\phi$ for $y_w$ and $y_l$ as $s_w$ and $s_l$, respectively.
Following Bradley-Terry model \cite{bradley_terry}, we establish: $p(y_w\succ y_l \mid x) = \frac{e^{s_w}}{e^{s_w}+e^{s_l}}$ and $p(y_w\prec y_l \mid x) = \frac{e^{s_l}}{e^{s_w}+e^{s_l}}$.
Then, the personalized label uncertainty is defined as:
\begin{equation}\epsilon_\phi = \frac{e^{s_l}}{(e^{s_w}+e^{s_l})},\end{equation}
representing the probability of label flipping (\ie $y_l$ winning $y_w$). 
And we adopt $p(y_w\succ y_l \mid x)=1-\epsilon_\phi$ as the target in Equation \eqref{eq:RosePO} .
In practice, we utilize a light yet efficient recommendation model --- SASRec \cite{sasrec}, as the approximation for the preference oracle $\phi$.
This allows for personalized preference predictions tailored to each data point, dynamically adjusting the optimization to account for uncertainty in user preferences.

%% file: chapters/exp.tex
\section{Experiments and Results}
\input{table/overall_performance_20-random-cans}
In this section, we conduct experiments on three datasets to demonstrate the effectiveness of our preference alignment framework, RosePO, which incorporates self-hard, semantic-similar, and popularity-aware rejected sampling, denoted as RosePO-h, RosePO-s, and RosePO-p in experiments, respectively. 
\begin{itemize}[leftmargin=*]
    \item From the perspective of \textit{helpfulness}, we evaluate the overall results of RosePO-h in Section \ref{sec:overall_performance}, along with detailed analyses on its two key components (\ie the optimization objective and the rejected sampling strategy) in Section \ref{sec:objective} and \ref{sec:sampling}.
    \item From \textit{harmlessness}, we illustrate how two rejected sampling approaches (\ie RosePO-s and RosePO-p) can mitigate semantic hallucination and popularity bias while preserving a satisfactory level of helpfulness comparable to SFT in Section \ref{sec:semantic} and \ref{sec:pop}.
\end{itemize}

\subsection{Experimental Settings}
\subsubsection{Datasets}
We conducted experiments on three distinct real-world datasets:
(1) MovieLens: a widely recognized recommendation dataset \footnote{https://grouplens.org/datasets/movielens/} comprising user ratings for various films;
(2) Steam: a dataset encompassing user ratings and reviews for games, which were gathered from a prominent video game digital distribution platform \footnote{https://store.steampowered.com/}; and (3) Goodreads: a dataset including user ratings and reviews of books, sourced from a website \footnote{https://www.goodreads.com/}, which is popular among readers for book recommendations.
To maintain stringent control over the scale of data, we have eliminated interactions with ratings less than 3 and 5 for the MovieLens and Goodreads datasets, respectively.
Furthermore, we have disregarded users with interactions less than 20 in Goodreads, and have sampled 30\% of the games and sequences for Steam. 
Subsequently, we split each dataset into train, validation, and test sets in an 8:1:1 ratio according to the sequence timestamp, thus preventing data leakage.
The sliding window to extract sequences is 11, with the last item in each sequence as the label for the next item.
Statistics of datasets are included in Appendix~\ref{app:dataset}.

\subsubsection{Implementation}
Following \cite{dros, llara}, our implementation of conventional recommender baselines utilizes the Adam optimizer with a learning rate of 1e-3, an embedding dimension of 64, and L2 regularization coefficient searched within the range of [1e-3, 1e-4, 1e-5, 1e-6, 1e-7]. 
For all LLM-based methods, we adapt their implementation to provide the ranking of candidate items.
Also, we train for only one epoch during each tuning stage, incorporating a warm-up strategy for the learning rate.
As for our method, we conduct experiments on 8 A800 NVIDIA RTX GPUs, and employ the widely-used Llama-3-8B-Instruct as the backbone, full tuning its parameters in both SFT and preference alignment stages.
All items in prompts are represented by their titles, serving as their textual features, and we use output probability distribution for item grounding.
Note that we perform self-hard rejected sampling strategy only on data points where SFT model made false predictions, while randomly sample when it predicts correctly.
We adopt an $\alpha$ of 0.2 following \cite{rpo}, and search $\beta$ within the sets [0.1, 0.2, 0.5, 1.0, 2.0].
Our batch sizes are 512, 256, and 256 for MovieLens, Steam, and Goodreads, respectively.
Considering randomness, we report the average outcomes of five runs using different random seeds.

\subsubsection{Evaluation}
Given the limited context length and slow inference speed of LLM, our method is more aptly suited for the fine-tuning phase of recommendation systems, focusing on ranking for a small number of items based on preferences.
As such, our primary experiments are conducted on 20 randomly selected non-interacted candidate items following \cite{llara}.
Correspondingly, we employ Hit Ratio (HR@K) and Normalized Discounted Cumulative Gain (NDCG@K) as metrics to evaluate recommendation performance, with K values of 1, 5, and 10.

\subsection{Performance Comparison} \label{sec:overall_performance}
\subsubsection{Baselines}
For traditional recommendation models, three commonly-used baselines are GRU4Rec \cite{gru4rec}, Caser \cite{caser}, and SASRec \cite{sasrec}, which are based on RNN, CNN and Attention Mechanism, respectively.
Regarding works that employ LLM in recommendation systems, we select three models from distinct categories --- LLM as an enhancer, LLM as a recommender with and without tuning – to serve as baselines:
(1) MoRec \cite{morec} enhances traditional recommendation systems by initialing its item embeddings with textual feature representations encoded by LLM. \footnote{For MoRec, we employ SASRec as the recommender backbone and BERT as the text encoder, in line with the officially provided implementation.}
(2) ChatRec \cite{chatrec} generates the recommendation list by prompting LLM without tuning it. \footnote{For ChatRec, we employ GPT-4 as a strong LLM backbone.}
(3) LLaRA \cite{llara} integrates traditional models into LLM using a projector and curriculum-tuning the LLM with prompts consisting of hybrid representations. \footnote{For LLaRA, we utilize Llama2-7B as the LLM backbone and employ LoRA for parameter-efficient fine-tuning, following the official implementation.}
Additionally, we provide results from three stages: the official pretrained checkpoint of Llama-3-8B-Instruct model, the checkpoint after recommendation SFT, and the final version after preference aligning by RosePO with self-hard rejected sampling, denoted as PT, SFT, and RosePO-h, respectively.

\subsubsection{Results}
As demonstrated in Table \ref{tab:overall_performance_20-random-cans}, RosePO-h surpasses all baseline methods (including conventional recommendation models and LLM4Rec works) across three datasets in terms of the evaluation metrics HR@1, HR@5, HR@10, NDCG@5, and NDCG@10.
This result highlights the exceptional efficacy of our proposed method.
Notably, the preference alignment stage exhibits improvements over two of its prior checkpoints (\ie PT and SFT).
Specifically, the relative enhancements of RosePO-h over SFT on the metric HR@1 are 8.11\%, 4.51\%, and 4.25\% for the Movielens, Goodreads, and Steam datasets, respectively.
This indicates that we addressed the inherent defects of the models in earlier stages, thereby enhancing the accuracy (\ie helpfulness) of the recommendations.

\subsection{Impact of Data-Adaptive Loss Design} \label{sec:objective}
To investigate the efficacy of our personalized smoothing loss function design, we conduct ablation studies to compare RosePO with various optimization objectives.

\subsubsection{Baselines}
We select DPO and its variants to serve as additional baselines for comparison. 
We have:
(1) \textbf{DPO} \cite{dpo}: offers a closed-form solution of the reward model in RLHF and optimizes the preference model offline.
(2) \textbf{IPO} \cite{ipo}: derives a general objective to bypass two approximations in DPO.
(3) \textbf{Conservative DPO} (cDPO) \cite{cdpo}: introduces a hyperparameter $\epsilon$ to model flip rate of noisy labels.
(4) \textbf{Robust DPO} (rDPO) \cite{rdpo}: designs an unbiased estimate of the original Binary Cross Entropy loss.
(5) \textbf{RPO} \cite{rpo}: includes a negative log-likelihood (NLL) loss term for the paired chose responses.
(6) \textbf{CPO} \cite{cpo}: employs sequence likelihood as a reward and trains jointly with an SFT objective.
(7) \textbf{SimPO} \cite{simpo}: utilizes the average log probability of a sequence as a reward to eliminate the need for a reference model, and introduces a target reward margin between the chosen and rejected responses.
(8) \textbf{Softmax DPO} (S-DPO) \cite{softmax_dpo}: introduces DPO into LLM4Rec by randomly sampling multiple negative items as rejected responses, to enhance the accuracy of recommendation. \footnote{We adapt S-DPO to the task setting of ranking item candidates, keeping the same implementation (including prompts and training) as ours, except for the softmax loss.}
Detailed objective formulations are listed in Table \ref{tab:objectives}.
\input{table/objectives}

\subsubsection{Results}
\input{table/ablation_xpo}
The experimental results of the ablation study\footnote{For Goodreads, IPO maintains anomalous gradients under various hyperparameters.} for objectives are presented in Table \ref{tab:ablation_xpo}.
It shows that RosePO surpasses the baselines across all five evaluation metrics on both Goodreads and Steam datasets.
In the case of the Movielens dataset, our method focuses on the fine-grained ranking of candidate items, rendering the HR@1, NDCG@5, and NDCG@10 metrics more crucial.
Thus, in these three metrics, our approach consistently outperforms the baselines on Movielens dataset.
In the context of Movielens, our method prioritizes the fine-grained ranking of candidate items, thus making the HR@1, NDCG@5, and NDCG@10 metrics of paramount importance. Our approach consistently surpasses the baselines across these three metrics.

This observation suggests that there indeed exists a certain level of noise within the preference labels constructed using our automated approach due to the incomplete user behavior in recommendation.
Consequently, this noise may be attributed to the ineffectiveness of optimization objectives, resulting in performance that is either inferior to our method or leads to a relative degradation in the model's performance when compared to the previous stage SFT model.
We incorporate personalized modeling for each preference label's flip rate, guided by a domain-specific model.
This customized objective, specifically tailored for recommendation, enhances the robustness of RosePO against noisy preference.

\subsection{Impact of Rejected Sampling Strategy} \label{sec:sampling}
\input{table/ablation_sampling}
In this section, we focus on the recommendation performance of different approaches to constructing rejected samples.

\subsubsection{Rejected Sampling Strategies}
We've compared four rejected sampling approaches as follows:
\begin{enumerate}[leftmargin=*]
    \item \textbf{Uniform}. This represents the most basic form of sampling --- uniformly sampling rejected items from all available items.
    \item \textbf{Semantic}. To mitigate the spurious correlation of LLMs relying heavily on semantics, we opt to select an item that is most semantically similar to the user's historical sequences.
    \item \textbf{Popular}. To enhance item-side fairness and encourage LLMs to recommend less mainstream or more niche items, we sample the rejected item based on the distribution of item popularity.
    \item \textbf{Self-hard}. To address the inherent shortcomings in the previous stage, we select cases where the SFT model gives incorrect outputs to serve as rejected responses. As for cases where the SFT model provided correct answers, we randomly sampled a negative item to serve as the rejected response.
\end{enumerate}

\subsubsection{Results}
As shown in Table \ref{tab:ablation_sampling}, RosePO-h (self-hard sampling) surpasses all other sampling strategies, particularly uniform sampling.
This highlights the significant advantage of the self-hard sampling approach in enhancing recommendation performance.
Other sampling strategies, such as uniform, semantic, and popular sampling, maintain comparable performance levels with the previous SFT model.

More specifically, on MovieLens, both semantic and popular sampling strategies show slightly lower performance in terms of metrics compared with the SFT model.
In the case of Steam, they are on par with the SFT model.
On Goodreads, both strategies outperform the SFT model across all five metrics.
This can be attributed to the fact that for these rejected sampling strategies, the preference alignment training might disrupt the SFT model's ability to make recommendations based on semantics and popularity.
Although semantics and popularity sometimes serve as shortcuts, they indeed enhance the accuracy of recommendation results.
Nevertheless, we still manage to maintain a certain degree of helpfulness in both RosePO-s and RosePO-p.

\subsection{Mitigating Semantic Hallucination} \label{sec:semantic}
To evaluate the effectiveness of RosePO-s in mitigating semantic hallucination, we examine two aspects: the metric of semantic hallucination and semantic-hard evaluation for ranking.

\subsubsection{Metric of Semantic Hallucination}
In order to examine the effectiveness of RosePO-s in mitigating semantic hallucination while achieving the performance displayed in Table \ref{tab:ablation_sampling}, we first need to quantify semantic hallucination.
We expect that the semantic similarity between the recommended items and the user's historical sequence will maintain a comparable similarity to the actual subsequent items that the user has interacted with.
Therefore, we define the semantic bias for the recommendation item $i_k$ as 
\begin{equation}
    \text{bias}_{\text{sem}}(u, i_k) = \frac{\text{similarity}_{\text{sem}}(u,i_k) - \text{similarity}_{\text{sem}}(u,i_{n+1})}{\text{similarity}_{\text{sem}}(u,i_{n+1})},
\end{equation}
with the semantic similarity \footnote{We adopt the MPNet as the semantic encoder.} given by:
\begin{equation}
    \text{similarity}_{\text{sem}}(u,i_k)= \frac{\sum_{t=1}^n \text{CosSim}([\text{emb}_k]), [\text{emb}_t])}{n}.
\end{equation}

As illustrated in the first line of Figure \ref{fig:bias}, we can observe that RosePO-s exhibits a leftward shift in the distribution of the semantic bias metric compared to the SFT model.
This indicates that, compared to the previous stage models, our approach has reduced its reliance on semantic information to a certain extent, thereby mitigating the semantic hallucination that may arise from it.

\subsubsection{Semantic-Hard Evaluation for Ranking}
\begin{figure*}
    \centering
    \begin{subfigure}[b]{0.28\textwidth}
        \includegraphics[width=\textwidth]{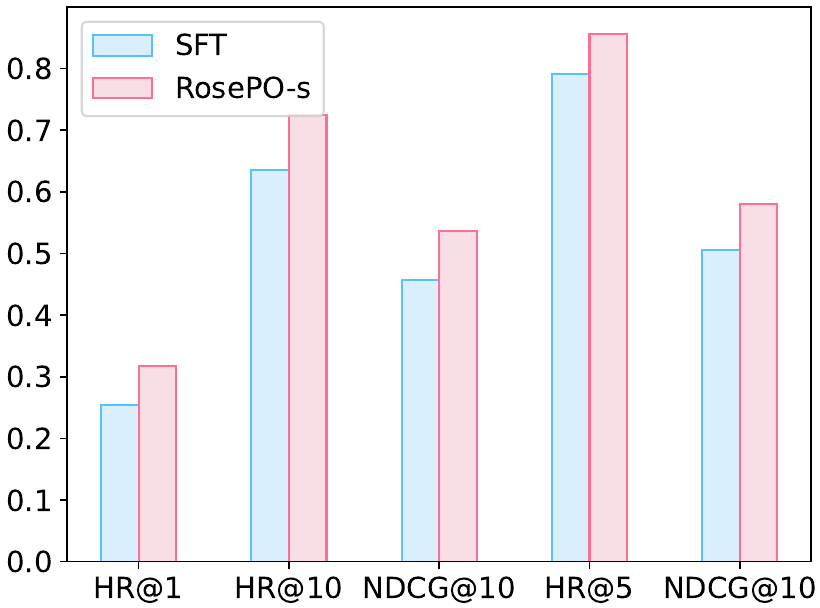}
        \caption{Movielens}
    \end{subfigure}
    \begin{subfigure}[b]{0.28\textwidth}
        \includegraphics[width=\textwidth]{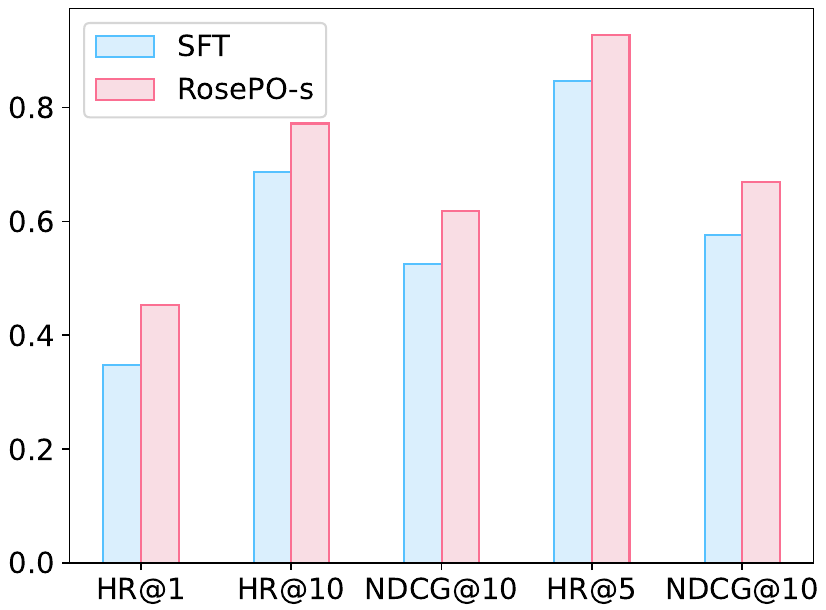}
        \caption{Goodreads}
    \end{subfigure}
    \begin{subfigure}[b]{0.28\textwidth}
        \includegraphics[width=\textwidth]{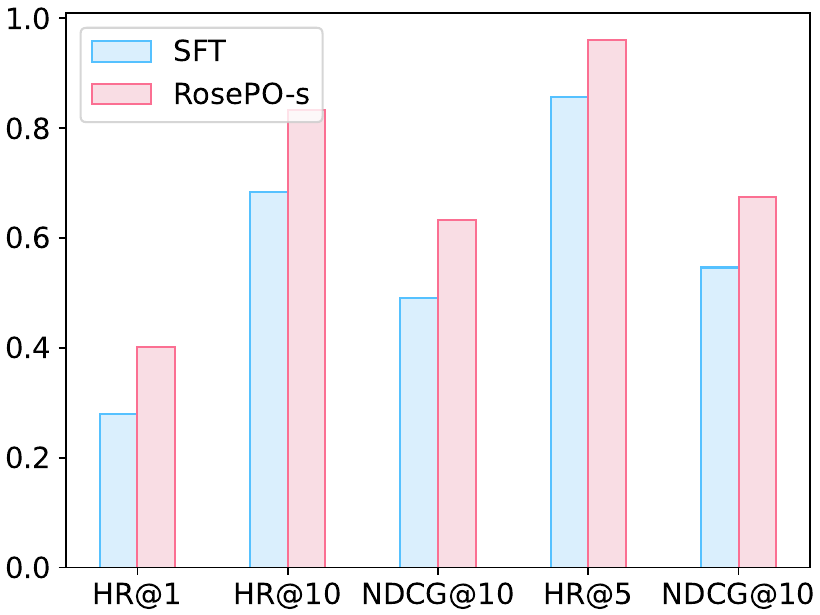}
        \caption{Steam}
    \end{subfigure}
    \vspace{-8pt}
    \caption{HR@1 of SFT and RosePO-s on ranking semantic-similar item candidates.}
    \vspace{-3pt}
    \label{fig:semantic-bias}
\end{figure*}
Moreover, we ask the recommender to rank item candidates sampled by confusing negatives that are semantically similar to the user's historical sequence, rather than the randomly sampled candidates.
In this challenging test setting, named semantic-hard setting, we observe a significant improvement in recommendation performance for RosePO-s compared to the SFT model as shown in Figure \ref{fig:semantic-bias}.
This demonstrates that RosePO can effectively teach the model to discern spurious correlations in semantics.

\subsection{Reducing Popularity Bias} \label{sec:pop}
In line with \cite{popularity_bias}, we define popularity bias as the difference between the popularity level of a recommendation item and the average popularity level of items in the user's historical sequence.
We formulate popularity bias as follows:
\begin{equation}
    \text{bias}_{\text{pop}}(u, i_t) = \text{LogPop}(i_t) - \sum_{k=1}^n \frac{\text{LogPop}(i_k)}{n},
\end{equation}
where $\text{LogPop}(\cdot)$ denotes the logarithmic popularity of an item.

\begin{figure*}
    \centering
    \begin{subfigure}[b]{0.25\textwidth}
        \includegraphics[width=\textwidth]{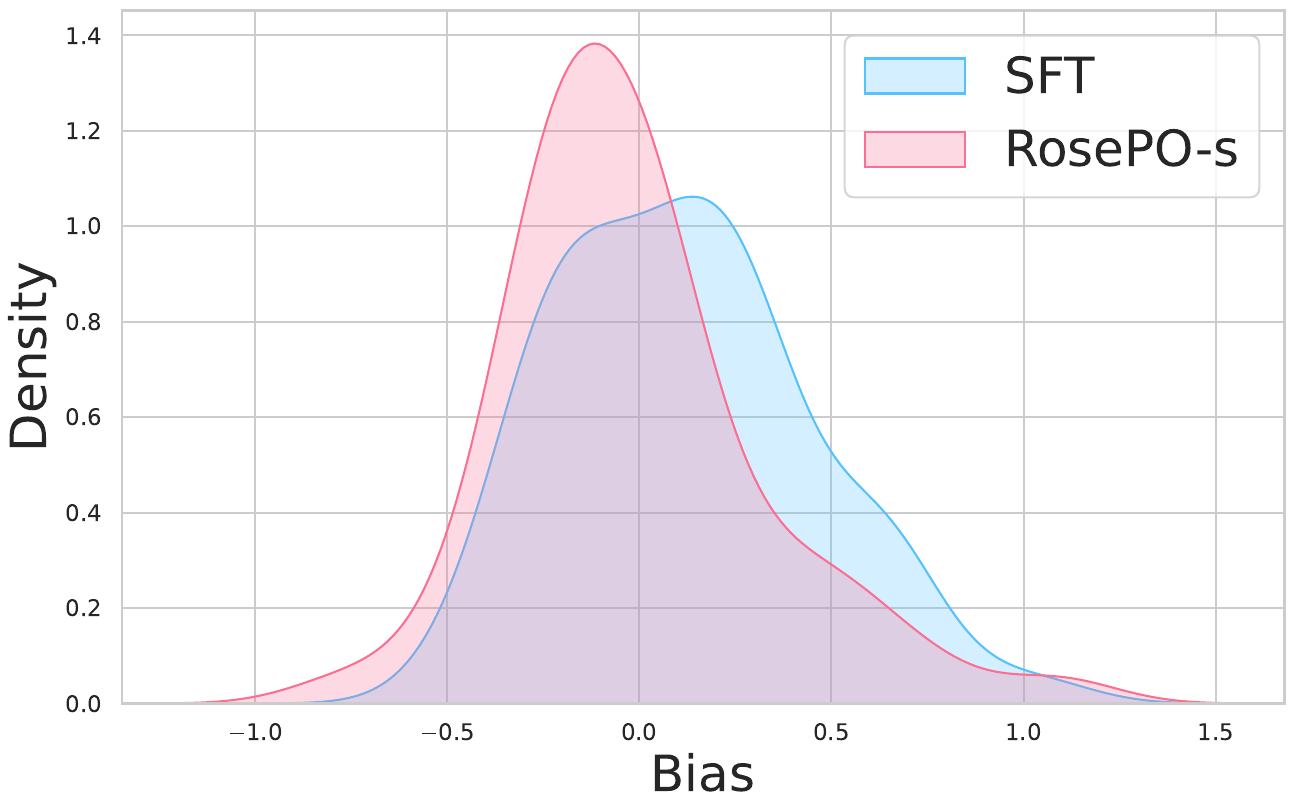}
        \caption{Semantic Bias on Movielens}
    \end{subfigure}
    \begin{subfigure}[b]{0.25\textwidth}
        \includegraphics[width=\textwidth]{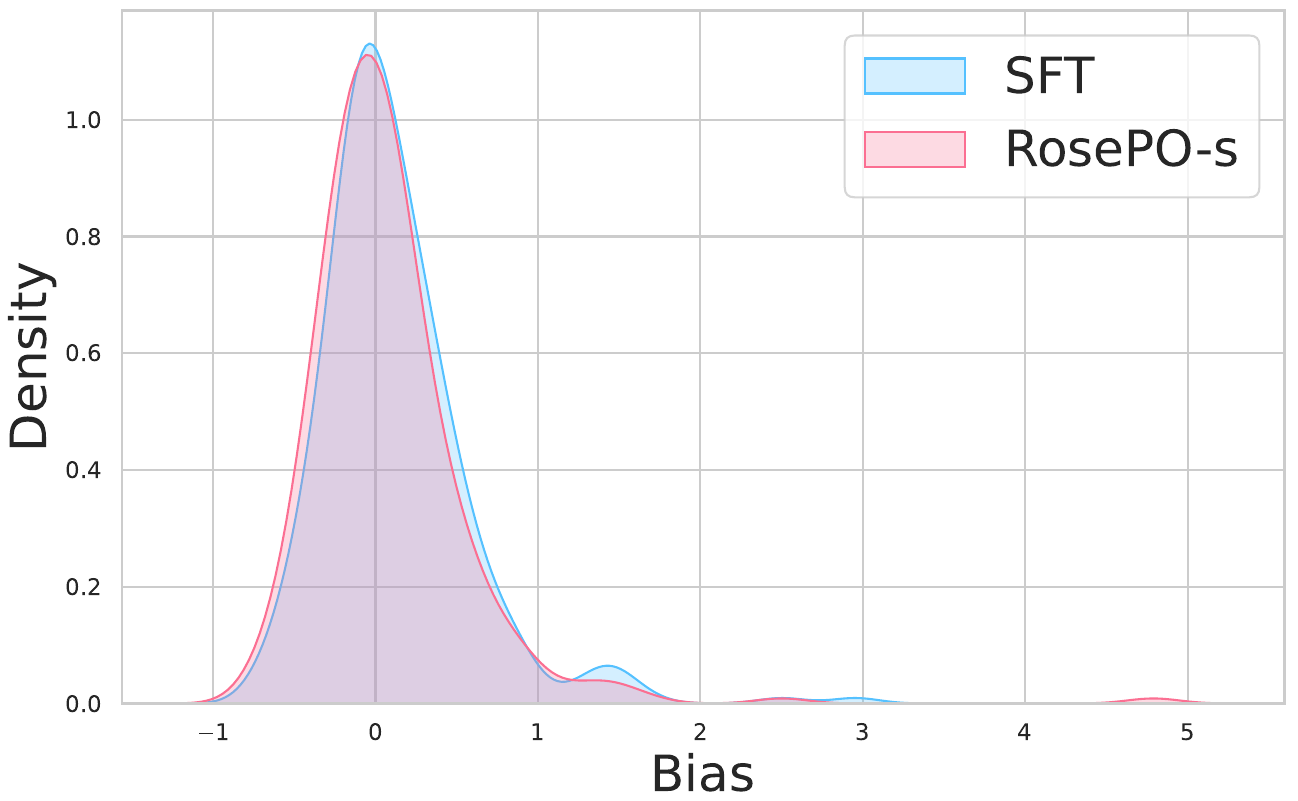}
        \caption{Semantic Bias on Goodreads}
    \end{subfigure}
    \begin{subfigure}[b]{0.25\textwidth}
        \includegraphics[width=\textwidth]{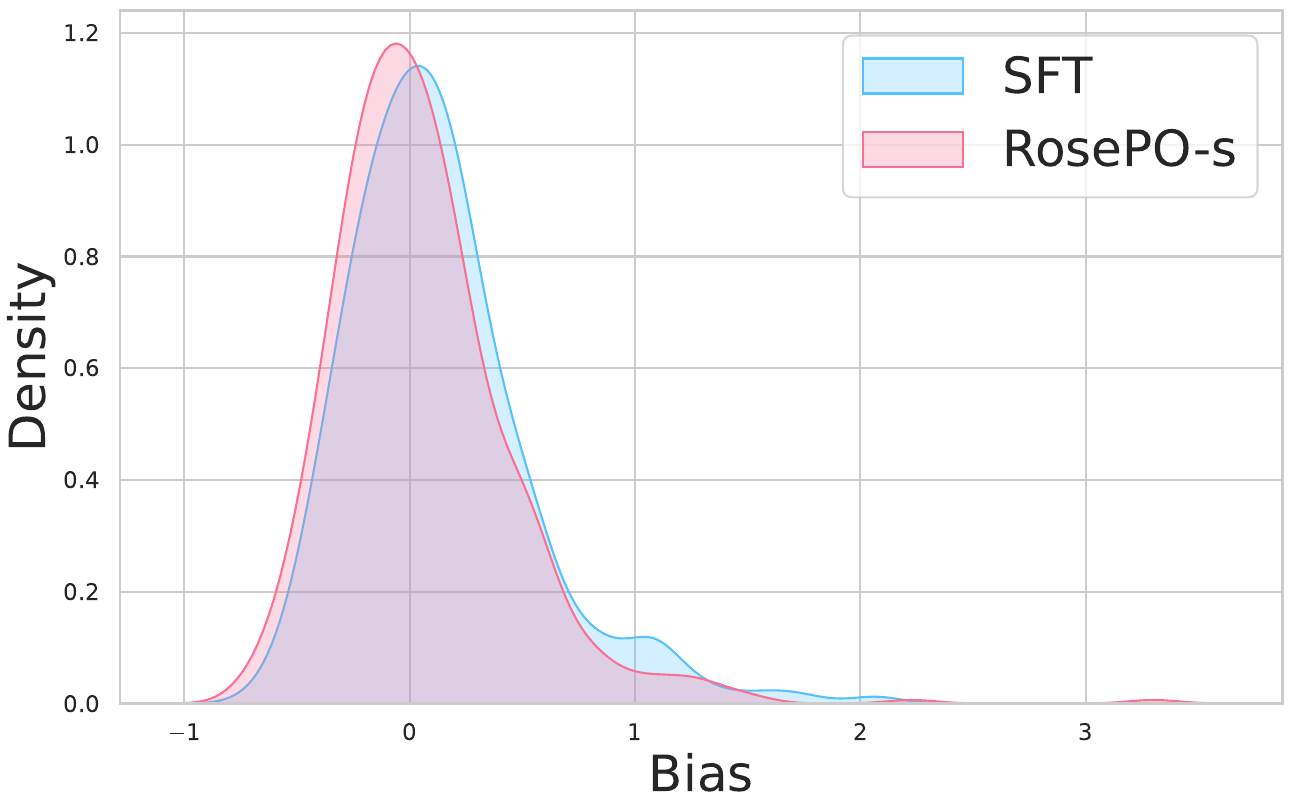}
        \caption{Semantic Bias on Steam}
    \end{subfigure}
    \newline
    \begin{subfigure}[b]{0.25\textwidth}
        \includegraphics[width=\textwidth]{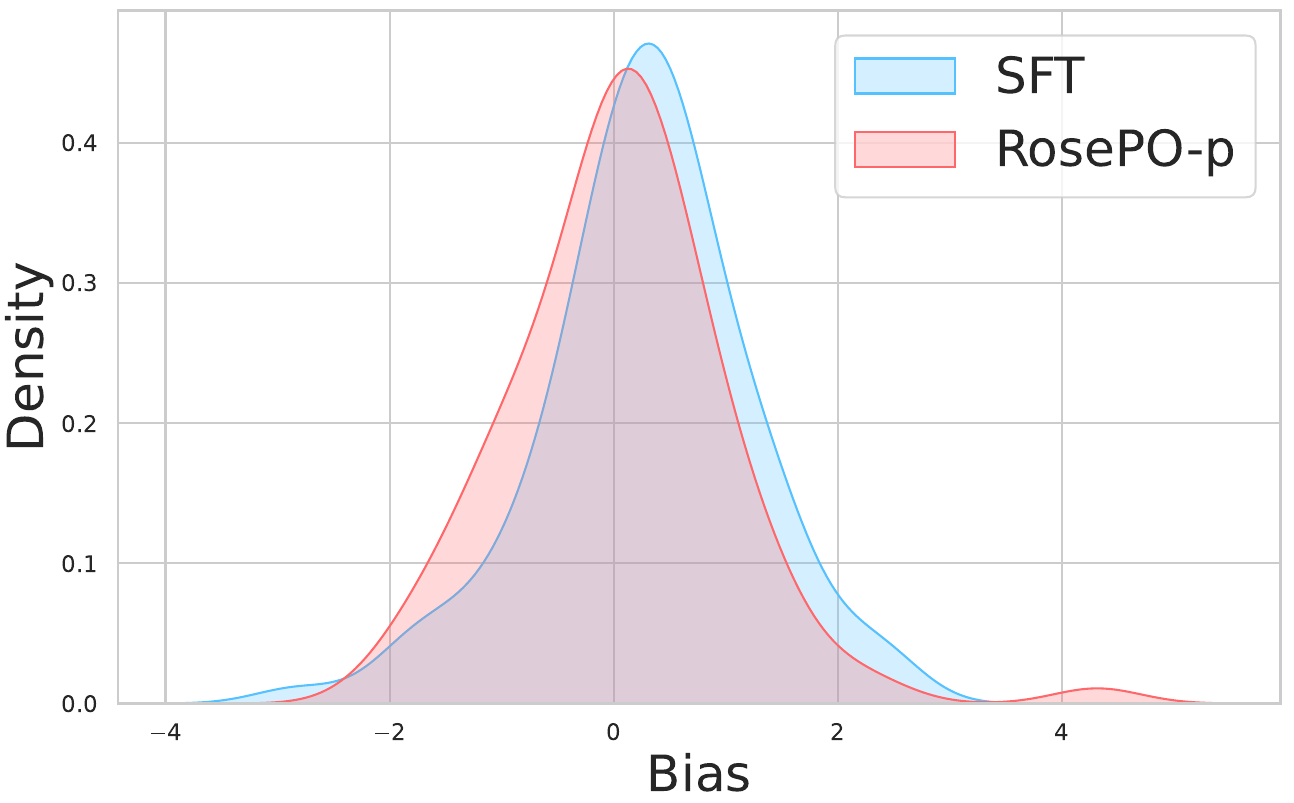}
        \caption{Popularity Bias on Movielens}
    \end{subfigure}
    \begin{subfigure}[b]{0.25\textwidth}
        \includegraphics[width=\textwidth]{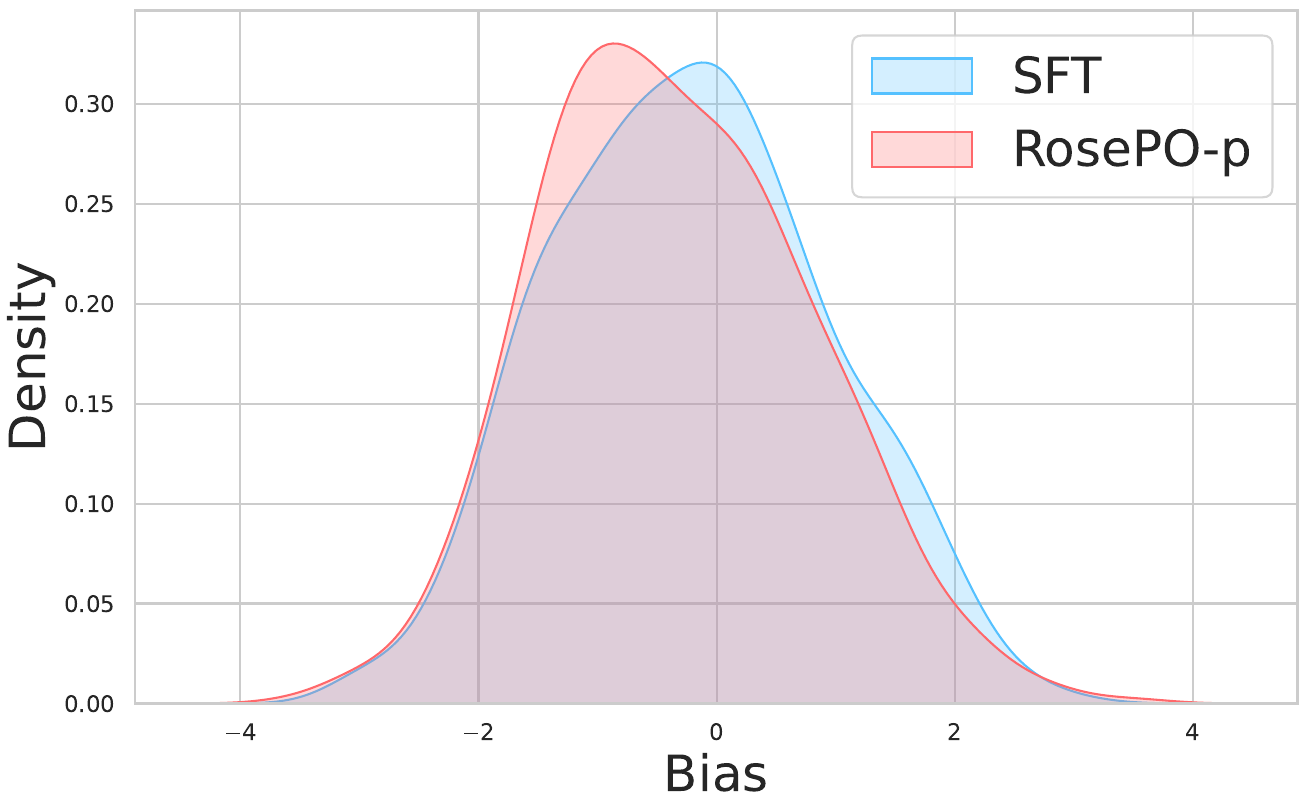}
        \caption{Popularity Bias on Goodreads}
    \end{subfigure}
    \begin{subfigure}[b]{0.25\textwidth}
        \includegraphics[width=\textwidth]{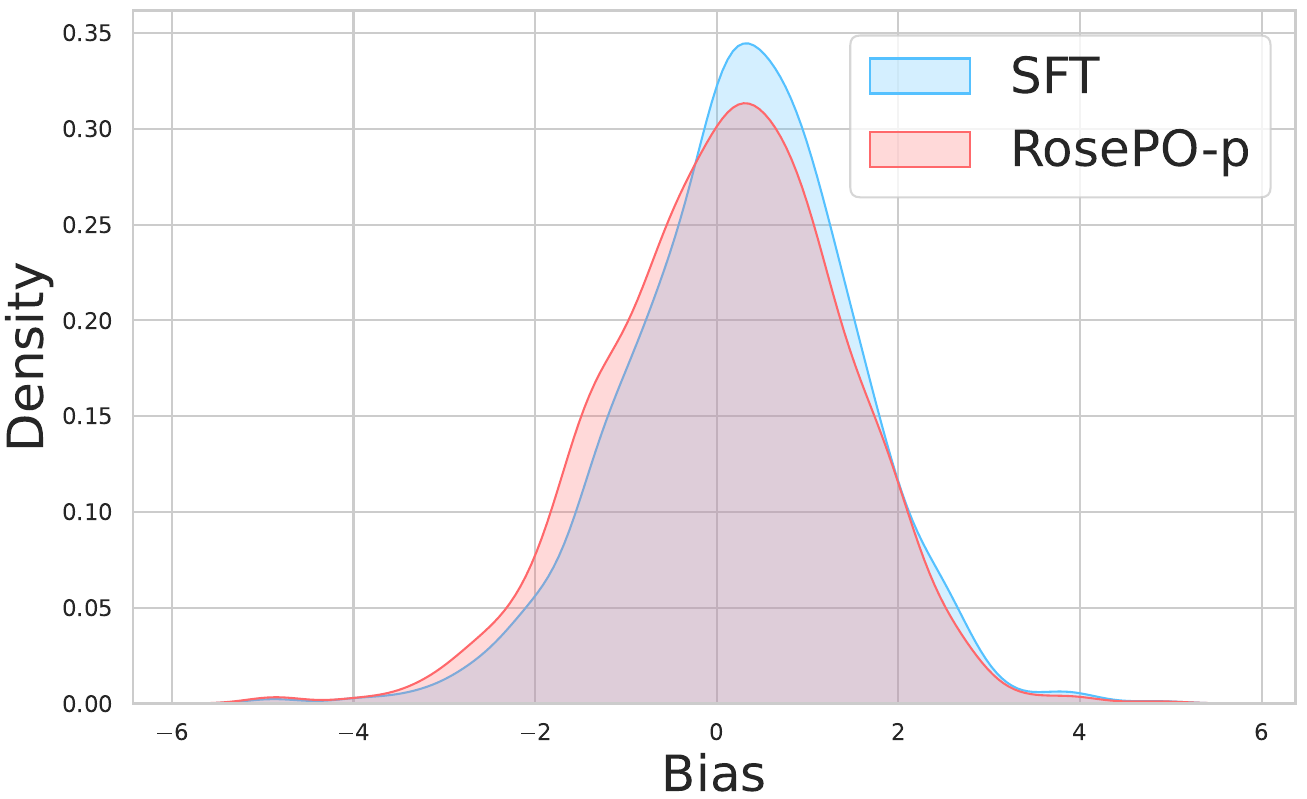}
        \caption{Popularity Bias on Steam}
    \end{subfigure}
    \vspace{-5pt}
    \caption{Distribution of semantic bias and popularity bias for SFT and RosePO (-s and -p).}
    \label{fig:bias}
    \vspace{-3pt}
\end{figure*}

As illustrated in the second line of Fig. \ref{fig:bias}, RosePO-p demonstrates a leftward shift in the distribution of the popularity bias metric when compared with the SFT model.
This suggests that RosePO-p is capable of recommending items that are less mainstream compared to those in the user's interaction history.
On one hand, this explores the user's interests, and on the other hand, it encourages the creation of niche items without compromising the user experience.

\subsection{Aligning with Multiple Preferences} \label{sec:multi-preference}
\input{table/multi-preference}
Going a step further, despite the potential conflicts between helpfulness and harmlessness \cite{3h}, we endeavor to address these three types of preferences concurrently using a hybrid sampling strategy, referred to as RosePO-m.
Specifically, for each $y_l$, the rejected sampling strategy is randomly selected from RosePO-h, -s, and -p.
We evaluate with six metrics considering both helpfulness (HR@1, HR@5, NDCG@5, and NDCG@10) and harmlessness ($\text{bias}_{\text{sem}}$ and $\text{bias}_\text{pop}$).
Table \ref{tab:hybrid} demonstrates that RosPO-m improves upon the SFT model across all evaluation metrics, moving towards a more HH recommendation service.

%% file: table/overall_performance_20-random-cans.tex
\begin{table*}[t]
\setlength{\abovecaptionskip}{0.05cm}
\setlength{\belowcaptionskip}{0cm}
\caption{Evaluation of helpfulness.}
\setlength{\tabcolsep}{2.7mm}{
\resizebox{\textwidth}{!}{
\begin{tabular}{l|ccccc|ccccc|ccccc}
\toprule
 & \multicolumn{5}{c|}{\textbf{Movielens}} & \multicolumn{5}{c|}{\textbf{Goodreads}} & \multicolumn{5}{c}{\textbf{Steam}} \\              
 \textbf{Model} & \textbf{HR@1} & \textbf{HR@5} & \textbf{N@5} & \textbf{HR@10} & \textbf{N@10} & \textbf{HR@1} & \textbf{HR@5} & \textbf{N@5} & \textbf{HR@10} & \textbf{N@10} & \textbf{HR@1} & \textbf{HR@5} & \textbf{N@5} & \textbf{HR@10} & \textbf{N@10} \\ \hline\hline
\textbf{GRU4Rec} & 0.2767 & 0.6200 & 0.4614 & 0.7100 & 0.4902 &  0.4048 & 0.7852 & 0.6087 & 0.9187 & 0.6523 & 0.3610 & 0.7324 & 0.5542 & 0.8886 & 0.6050 \\
\textbf{Caser} & 0.3783 & 0.6383 & 0.5143 & 0.7267 & 0.5430 & 0.4161 & 0.8128 & 0.6284 & 0.9187 & 0.6630 & 0.3810 & 0.7587 & 0.5803 & 0.8879 & 0.6225 \\
\textbf{SASRec} & 0.2817 & 0.6183 & 0.4598 & 0.7184 & 0.4928 & 0.4102 & 0.7770 & 0.6067 & 0.9032 & 0.6475 & 0.3927 & 0.74856 & 0.5817 & 0.8699 & 0.6211 \\
\midrule
\textbf{MoRec} & 0.2050 & 0.5083 & 0.3638 & 0.6716 & 0.4168 & 0.3152 & 0.7025 & 0.5147 & 0.8635 & 0.5670 & 0.3942 & 0.7617 & 0.5918 & 0.8857 & 0.6323 \\
\textbf{Chat-REC} & 0.2105 & 0.6842 & 0.4534 & 0.8211 & 0.4978 & 0.3694 & 0.7088 & 0.5412 & 0.8336 & 0.5814 & 0.4012 & 0.7306 & 0.5785 & 0.8370 & 0.6128 \\
\textbf{LLaRA} & 0.4379 & 0.7895 & 0.6285 & 0.8884 & 0.6602 & 0.5158 & 0.8060 & 0.6747 & 0.8859 & 0.7005 & 0.4927 & 0.8749 & 0.6988 & 0.9511 & 0.7236 \\
\midrule
\textbf{PT} & 0.0968 & 0.3074 & 0.2032 & 0.5347 & 0.2753 & 0.1191 & 0.3418 & 0.2320 & 0.5820 & 0.3090 & 0.0868 & 0.3506 & 0.2214 & 0.5793 & 0.2947 \\
\textbf{SFT} & 0.4674 & 0.8800 & 0.6865 & \textbf{0.9537} & 0.7111 & 0.5165 & 0.8692 & 0.7053 & 0.9667 & 0.7373 & 0.5105 & 0.8788 & 0.7100 & 0.9548 & 0.7348 \\
\cellcolor{gray!16}\textbf{RosePO-h} & \cellcolor{gray!16}\textbf{0.5053} & \cellcolor{gray!16}\textbf{0.8926} & \cellcolor{gray!16}\textbf{0.7144} & \cellcolor{gray!16}\textbf{0.9537} & \cellcolor{gray!16}\textbf{0.7348} & \cellcolor{gray!16}\textbf{0.5398} &\cellcolor{gray!16}\textbf{0.8805} &\cellcolor{gray!16}\textbf{0.7221} &\cellcolor{gray!16}\textbf{0.9717} &\cellcolor{gray!16}\textbf{0.7520} &\cellcolor{gray!16}\textbf{0.5322}  &\cellcolor{gray!16}\textbf{0.8836}  &\cellcolor{gray!16}\textbf{0.7239}  &\cellcolor{gray!16}\textbf{0.9619}  &\cellcolor{gray!16}\textbf{0.7494} \\ \bottomrule
\end{tabular}
}}
\label{tab:overall_performance_20-random-cans}
\vspace{-8pt}
\end{table*}

%% file: table/objectives.tex
\begin{table}[t]
\setlength{\abovecaptionskip}{0.05cm}
\setlength{\belowcaptionskip}{0cm}
\caption{Optimization Objectives for Preference Alignment.}
\setlength{\tabcolsep}{2.7mm}{
\resizebox{8.5cm}{!}{
\begin{tabular}{ll}
\toprule
\textbf{Method} & \textbf{Objective}  \\ \midrule
\midrule 
DPO \cite{dpo} & $-\log \sigma \left( \beta \log \frac{\pi_\theta(y_w|x)}{\pi_{\text{ref}}(y_w|x)} - \beta \log \frac{\pi_\theta(y_l|x)}{\pi_{\text{ref}}(y_l|x)}\right)$ \\ \midrule 
IPO~\cite{ipo} & $ \left( \log \frac{\pi_\theta(y_w|x)}{\pi_{\text{ref}}(y_w|x)} - \log \frac{\pi_\theta(y_l|x)}{\pi_{\text{ref}}(y_l|x)} - \frac{1}{2\tau} \right)^2 $ \\  \midrule 
cDPO \cite{cdpo} & $-(1-\epsilon)\left( \beta \log \frac{\pi_\theta(y_w|x)}{\pi_{\text{ref}}(y_w|x)} - \beta \log \frac{\pi_\theta(y_l|x)}{\pi_{\text{ref}}(y_l|x)}\right) $ \\
& $- \epsilon \left(\beta \log \frac{\pi_\theta(y_l|x)}{\pi_{\text{ref}}(y_l|x)} - \beta \log \frac{\pi_\theta(y_w|x)}{\pi_{\text{ref}}(y_w|x)} \right)$ \\ 
\midrule
rDPO \cite{rdpo} & $ \frac{-(1-\epsilon)\left( \beta \log \frac{\pi_\theta(y_w|x)}{\pi_{\text{ref}}(y_w|x)} - \beta \log \frac{\pi_\theta(y_l|x)}{\pi_{\text{ref}}(y_l|x)}\right) + \epsilon \left(\beta \log \frac{\pi_\theta(y_l|x)}{\pi_{\text{ref}}(y_l|x)} - \beta \log \frac{\pi_\theta(y_w|x)}{\pi_{\text{ref}}(y_w|x)} \right)}{1-2\epsilon}$\\
\midrule
RPO \cite{rpo} & $-\log \sigma \left( \beta \log \frac{\pi_\theta(y_w|x)}{\pi_{\text{ref}}(y_w|x)} - \beta \log \frac{\pi_\theta(y_l|x)}{\pi_{\text{ref}}(y_l|x)}\right) -\alpha \frac{\pi_\theta(y_w|x)}{\mid y_w \mid}$ \\ \midrule
CPO \cite{cpo} &  $-\log \sigma  \left(\beta \log \pi_\theta(y_w|x) - \beta \log \pi_\theta(y_l|x) \right) - \lambda \log \pi_\theta (y_w|x)$ \\ \midrule
SimPO \cite{simpo} & $-\log \sigma  \left( \frac{\beta}{|y_w|} \log \pi_\theta(y_w|x) - \frac{\beta}{|y_l|} \log \pi_\theta(y_l|x) - \gamma \right)$ \\ \midrule
S-DPO \cite{softmax_dpo} & $-\log \sigma  \left( -\log\sum_{y_l\in\mathcal{Y}_l}\exp\left(\beta \log \frac{\pi_\theta(y_w|x)}{\pi_{\text{ref}}(y_w|x)} - \beta \log \frac{\pi_\theta(y_l|x)}{\pi_{\text{ref}}(y_l|x)}\right) \right)$ \\ \midrule
\textbf{RosePO} \cite{cdpo} & $-(1-\epsilon_\phi)\left( \beta \log \frac{\pi_\theta(y_w|x)}{\pi_{\text{ref}}(y_w|x)} - \beta \log \frac{\pi_\theta(y_l|x)}{\pi_{\text{ref}}(y_l|x)}\right) $ \\
& $- \epsilon_\phi \left(\beta \log \frac{\pi_\theta(y_l|x)}{\pi_{\text{ref}}(y_l|x)} - \beta \log \frac{\pi_\theta(y_w|x)}{\pi_{\text{ref}}(y_w|x)} \right)$ \\ 
\bottomrule
\end{tabular}
}}
\label{tab:objectives}
\vspace{-8pt}
\end{table}

%% file: table/ablation_xpo.tex
\begin{table*}[t]
\setlength{\abovecaptionskip}{0.05cm}
\setlength{\belowcaptionskip}{0cm}
\caption{Ablations for optimization objectives.}
\setlength{\tabcolsep}{2.7mm}{
\resizebox{\textwidth}{!}{
\begin{tabular}{l|ccccc|ccccc|ccccc}
\toprule
 & \multicolumn{5}{c|}{\textbf{Movielens}} & \multicolumn{5}{c|}{\textbf{Goodreads}} & \multicolumn{5}{c}{\textbf{Steam}} \\              
 \textbf{Model} & \textbf{HR@1} & \textbf{HR@5} & \textbf{N@5} & \textbf{HR@10} & \textbf{N@10} & \textbf{HR@1} & \textbf{HR@5} & \textbf{N@5} & \textbf{HR@10} & \textbf{N@10} & \textbf{HR@1} & \textbf{HR@5} & \textbf{N@5} & \textbf{HR@10} & \textbf{N@10} \\ \hline\hline
\textbf{DPO} & 0.4484 & 0.8821 & 0.6804 & 0.9600 & 0.7061 & 0.5092 & 0.8699 & 0.7003 & 0.9634 & 0.7309 & 0.4669 & 0.8528 & 0.6756 & 0.9477 & 0.7068 \\
$\textbf{IPO}^8$ & 0.4716 & \textbf{0.8926} & 0.6999 & 0.9537 & 0.7202 & - & - & - & - & - & 0.5020 & 0.8644 & 0.6978 & 0.9459 & 0.7242 \\
\textbf{cDPO} & 0.4821 & 0.8758 & 0.6974 &  0.9516 & 0.7222& 0.5131 & 0.8666 & 0.7004 & 0.9637 & 0.7320 & 0.4744 & 0.8589 & 0.6826 & 0.9489 & 0.7120 \\
\textbf{rDPO} & 0.4505 & 0.8863 & 0.6872 & \textbf{0.9621} & 0.7120 & 0.5095 & 0.8669 & 0.6992 & 0.9637 & 0.7309 & 0.4761 & 0.8541 & 0.6809 & 0.9489 & 0.7118 \\
\textbf{RPO} & 0.5011 & 0.8800 & 0.7054 & 0.9495 & 0.7291 & 0.5344 & 0.8779 & 0.7182 & 0.9700 & 0.7484 & 0.5293 & 0.8794 & 0.7205 & 0.9612 & 0.7472 \\
\textbf{CPO} & 0.4547 & 0.8526 & 0.6688 &  0.9537 & 0.7026 & 0.5298 & 0.8725 & 0.7113 & 0.9681 & 0.7428 & 0.5051 & 0.8712 & 0.7049& 0.9558& 0.7325 \\
\textbf{SimPO} & 0.3389 & 0.8063 & 0.5833 &  0.9453 & 0.6294 & 0.4369 & 0.8316 & 0.6452 & 0.9524 & 0.6848 & 0.3211 & 0.7909 & 0.5684 & 0.9170 & 0.6096 \\
\textbf{S-DPO} & 0.4632 & 0.8505 & 0.6749 & 0.9011 & 0.6915 & 0.5188 & 0.7890 & 0.6648 & 0.8839 & 0.6953 & 0.5101 & 0.8250 & 0.6830 & 0.9177 & 0.7132  \\
\midrule
\cellcolor{gray!16}\textbf{RosePO-h} & \cellcolor{gray!16}\textbf{0.5053} & \cellcolor{gray!16}0.8926 & \cellcolor{gray!16}\textbf{0.7144} & \cellcolor{gray!16}0.9537 & \cellcolor{gray!16}\textbf{0.7348} & \cellcolor{gray!16}\textbf{0.5398} &\cellcolor{gray!16}\textbf{0.8805} &\cellcolor{gray!16}\textbf{0.7221} &\cellcolor{gray!16}\textbf{0.9717} &\cellcolor{gray!16}\textbf{0.7520} &\cellcolor{gray!16}\textbf{0.5322}  &\cellcolor{gray!16}\textbf{0.8836}  &\cellcolor{gray!16}\textbf{0.7239}  &\cellcolor{gray!16}\textbf{0.9619}  &\cellcolor{gray!16}\textbf{0.7494} \\ \bottomrule
\end{tabular}
}}
\label{tab:ablation_xpo}
\end{table*}

%% file: table/ablation_sampling.tex
\begin{table*}[t]
\vspace{-0.2cm}
\setlength{\abovecaptionskip}{0.05cm}
\setlength{\belowcaptionskip}{0cm}
\caption{Ablations for rejected sampling strategies.}
\setlength{\tabcolsep}{2.7mm}{
\resizebox{\textwidth}{!}{
\begin{tabular}{l|ccccc|ccccc|ccccc}
\toprule
 & \multicolumn{5}{c|}{\textbf{Movielens}} & \multicolumn{5}{c|}{\textbf{Goodreads}} & \multicolumn{5}{c}{\textbf{Steam}} \\              
 \textbf{Model} & \textbf{HR@1} & \textbf{HR@5} & \textbf{N@5} & \textbf{HR@10} & \textbf{N@10} & \textbf{HR@1} & \textbf{HR@5} & \textbf{N@5} & \textbf{HR@10} & \textbf{N@10} & \textbf{HR@1} & \textbf{HR@5} & \textbf{N@5} & \textbf{HR@10} & \textbf{N@10} \\ \hline\hline
\textbf{uniform} & 0.4632 & 0.8695 & 0.6850 & 0.9305 & 0.7056 & 0.5285 & 0.8799 & 0.7171 & 0.9671 & 0.7458 & 0.5283 & 0.8811 & 0.7218 & 0.9592 & 0.7472 \\
\textbf{semantic} & 0.4400 & 0.8863 & 0.6843 & 0.9474 & 0.7042 & 0.5314 & 0.8819 & 0.7206 & 0.9661 & 0.7482 & 0.5121 & 0.8836 & 0.7143 & 0.9563 & 0.7379 \\
\textbf{popular} & 0.4758 & 0.8842 & 0.6981 & 0.9495 & 0.7196 & 0.5328 & 0.8812 & 0.7198 & 0.9657 & 0.7477 & 0.5261 & 0.8836 & 0.7219 & 0.9599 & 0.7467 \\
\textbf{self-hard} & \textbf{0.5053} & \textbf{0.8926} & \textbf{0.7144} & \textbf{0.9537} & \textbf{0.7348} & \textbf{0.5398} &\textbf{0.8805} &\textbf{0.7221} &\textbf{0.9717} &\textbf{0.7520} &\textbf{0.5322}  &\textbf{0.8836}  &\textbf{0.7239}  &\textbf{0.9619}  &\textbf{0.7494} \\ \bottomrule
\end{tabular}
}}
\label{tab:ablation_sampling}
\end{table*}

%% file: table/multi-preference.tex
\begin{table}[t]
\setlength{\abovecaptionskip}{0.05cm}
\setlength{\belowcaptionskip}{0cm}
\caption{HH evaluation for hybrid rejected sampling.}
\setlength{\tabcolsep}{2.7mm}{
\resizebox{0.47\textwidth}{!}{
\begin{tabular}{l|cccc|cc}
\toprule
& \multicolumn{4}{c|}{\textbf{Helpfulness}} & \multicolumn{2}{c}{\textbf{Harmlessness}}\\
\textbf{Model} & \textbf{HR@1}$\uparrow$ & \textbf{HR@5}$\uparrow$ &  \textbf{N@5}$\uparrow$ & \textbf{N@10}$\uparrow$ &\textbf{Sem. Bias}$\downarrow$ & \textbf{Pop. Bias}$\downarrow$ \\ \hline\hline
\multicolumn{7}{l}{\cellcolor{gray!16}\textbf{Movielens}}\\    
\midrule
\textbf{SFT} & 0.4674 & 0.8800 & 0.6865 & 0.7111 & 0.0850 & 0.1859 \\
\textbf{RosePO-m} & \textbf{0.4737} & \textbf{0.8863} & \textbf{0.6959} & \textbf{0.7166} & \textbf{0.0424} & \textbf{0.1836} \\
\hline\hline
\multicolumn{7}{l}{\cellcolor{gray!16}\textbf{Goodreads}}\\    
\midrule
\textbf{SFT} & 0.5165 & 0.8692 & 0.7053 & 0.7373 & 0.1099 & -0.2583 \\
\textbf{RosePO-m} & \textbf{0.5401} & \textbf{0.8819} & \textbf{0.7225} & \textbf{0.7502} & \textbf{0.0774} &  \textbf{-0.3300} \\
\midrule
\multicolumn{7}{l}{\cellcolor{gray!16}\textbf{Steam}}\\        \hline\hline
\textbf{SFT} & 0.5105 & 0.8788 & 0.7100 & 0.7348 & 0.1143 & 0.2458 \\
\textbf{RosePO-m} & \textbf{0.5261} & \textbf{0.8858} & \textbf{0.7231} & \textbf{0.7476} & \textbf{0.0866} & \textbf{0.1743} \\
\bottomrule
\end{tabular}
}}
\label{tab:hybrid}
\vspace{-10pt}
\end{table}

%% file: chapters/conclusion.tex
\section{Conclusion}
In this paper, we underscore the necessity of the preference alignment stage in developing a helpful and harmless LLM-based recommender.
To curate data with specific preferences and handle different levels of noise in preference labels, we introduce Recommendation with smoothing personalized Preference Optimization --- RosePO.
Conclusively, we made following contributions:
(1) Taking three preferences as examples, we design rejected sampling strategies to overcome model limitations in providing more accurate recommendations (\ie helpfulness), as well as mitigating semantic hallucination and popularity bias (\ie harmlessness).
(2) We incorporate a personalized smoothing factor into the optimization objective, leading to more robust preference alignment.
(3) Thorough experimental analysis demonstrates that RosePO is effective in aligning the LLM-based recommender with specific preferences, and the two key components of RosePO are essential.

While we recognize the positive impact LLMs have in recommendations, we also cannot ignore the realities revealed by the flip side of the coin.
This paper emphasizes the importance of human-centric values in LLM-based recommenders, paving the way for a more helpful and harmless service that better satisfies user needs and expectations.

%% file: chapters/appendix.tex
\appendix
\section{Datasets}
\label{app:dataset}
\input{table/dataset}

\section{Limitation}
Despite our RosePO achieving promising results in HH recommendation, we must acknowledge several limitations:
\label{app:limitation}

(1) Applied stage.
We only focus on ranking item candidates as in \cite{llara, uncovering_chatgpt4rec, is_chatgpt_a_good_rs}.
To explore the generalization capabilities of our RosePO to other recommendation stages, we demonstrate RosePO-h's helpfulness in ranking all items in Appendix \ref{sec:all-ranking}.
Future work should consider preference alignment in multiple stages (\eg recall, pre-ranking, ranking, re-ranking \cite{mrs}) of a recommender system.

(2) Data scale.
Due to limited computational resources, we conduct experiments only on relatively small-scale datasets (\cf Table~\ref{tab:data_statis}).
Nonetheless, it requires approximately 24 A800 GPU hours of training for the complete preference dataset.
Additionally, we investigate the impact of varying quantities of preference data on the recommendation performance in Appendix \ref{sec:scale}.
To scale up to larger datasets, future work should explore improvements on efficiency.

(3) Preference types.
We present three examples of helpfulness and harmlessness to illustrate the handling of a specific preference and demonstrate the integration of these preferences.
In future work, a wider range of preferences can be considered, such as promoting newly emerged items.

(4) Negative sampling.
Our work injects preferences into the negative space, with the core intent of constructing informative negative samples and further addressing the issue of false negatives.
This is closely related to the work on negative sampling \cite{sgl, ns_survey}.
Future research should focus on the problems associated with introducing LLM to recommendation and design more refined negative sampling approaches.

Our future work will be dedicated to addressing these limitations and improving the recommendation service.

\section{Performance Comparison on All-Ranking.} \label{sec:all-ranking}
In this section, we demonstrate the performance of RosePO-h to rank all items on three datasets (Movielens, Goodreads, and Steam).

\vspace{5pt}
\noindent
\textbf{Baselines.}
We choose both traditional sequential recommendation models (GRU4Rec \cite{gru4rec}, Caser \cite{caser}, SASRec \cite{sasrec}) and LLM4Rec baselines (BIGRec \cite{bigrec}) for comparison.
BIGRec, a representative LLM4Rec work on all-ranking, grounds the generated tokens to actual items in a bi-step manner, incorporating both the semantic similarity in hidden states and popularity information.

\vspace{5pt}
\noindent
\textbf{Implementation of Item Grounding.}
For BIGRec, we search the balanced hyperparameter $\gamma$ in [0.0, 0.2, 0.4, 0.6, 0.8, 1.0, 2.0, 5.0, 10.0, 20.0, 50.0, 100.0] to explore a better integration.
For RosePO, we conduct the beam search to adapt it into a generative style.

\vspace{5pt}
\noindent
\textbf{Evaluation Metrics.}
Considering the item sizes of three datasets, we compute HR@20 and NDCG@20 as evaluation metrics.

\vspace{5pt}
\noindent
\textbf{Results.}
We can observe from Table \ref{tab:overall_all-ranking_performance} that RosePO-h not only outperforms all the baselines but also improves the recommendation performance of the previous stage (SFT model).
This demonstrates that, when compared to other methods focusing on helpful recommendations, our RosePO-h method exhibits superior advantages.

\input{table/all-ranking_performance}

\section{Data Scale in Preference Learning} \label{sec:scale}
\begin{figure}[t]
\centering
\includegraphics[width=0.35\textwidth]{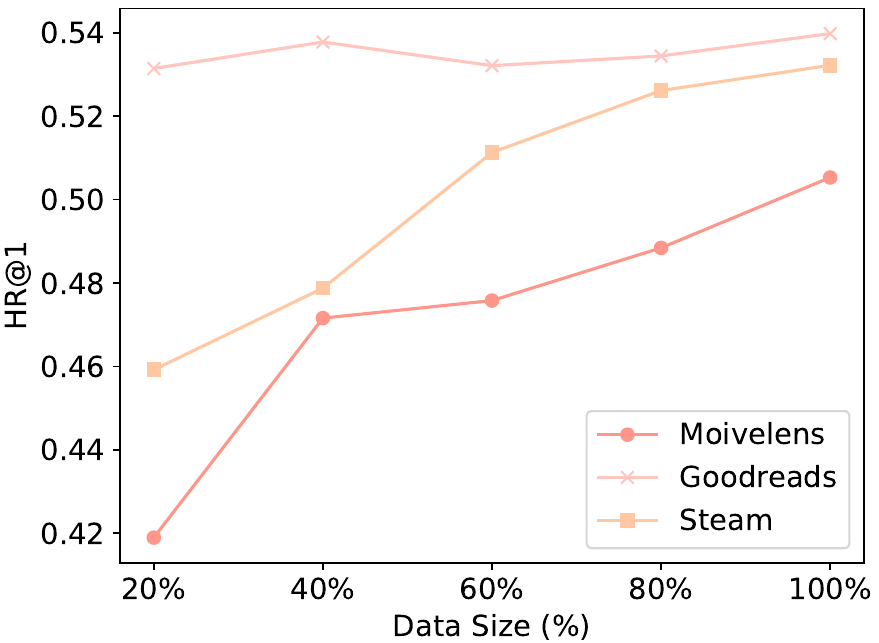}
\caption{The performance of RosePO-h for different scales of samples.}
\label{fig:scale}
\vspace{-5pt}
\end{figure}
Figure \ref{fig:scale} presents the HR@1 of RosePO-h across three datasets for different percentages of data. We can observe from it that, as the amount of data increases, the overall recommendation effectiveness demonstrates an upward trend.
Furthermore, only 60\% of the data is required to surpass the performance of SFT across all three datasets on the metric of HR@1. Especially, a mere 20\% of the data is sufficient to outperform SFT on the Goodreads dataset on HR@1. The trend of continuous improvement also implies that more preference learning samples are very likely to bring in further enhancement.

%% file: table/dataset.tex
\begin{table}[h]
\centering
\caption{Statistics of Datasets. }
\vspace{-8pt}
\label{tab:data_statis}
\begin{small}
\begin{tabular}{lrrr}
\toprule
Dataset& MovieLens & Goodreads & Steam  \\
\midrule
\# Sequence & 943 & 6,031 & 11,938 \\ 
\# Item & 1,682 & 4,550 & 3,581 \\ 
\# Interaction & 100,000 & 220,100 & 274,726 \\ 
\bottomrule
\end{tabular}
\end{small}
\vspace{-8pt}
\end{table}

%% file: table/all-ranking_performance.tex
\begin{table}[t]
\setlength{\abovecaptionskip}{0.05cm}
\setlength{\belowcaptionskip}{0cm}
\caption{Performance comparison for all-ranking.}
\setlength{\tabcolsep}{2.7mm}{
\resizebox{0.47\textwidth}{!}{
\begin{tabular}{l|cc|cc|cc}
\toprule
 & \multicolumn{2}{c|}{\textbf{Movielens}} & \multicolumn{2}{c|}{\textbf{Goodreads}} & \multicolumn{2}{c}{\textbf{Steam}} \\              
 \textbf{Model} & \textbf{HR@20} & \textbf{N@20} &  \textbf{HR@20} & \textbf{N@20} & \textbf{HR@20} & \textbf{N@20} \\ \hline\hline
\textbf{GRU4Rec} & 0.1667 & 0.0527 & 0.1143 & 0.0478 & 0.1308 & 0.0505 \\
\textbf{Caser} & 0.1550 & 0.0668 & 0.1238 & 0.0502 & 0.1260 & 0.0501 \\
\textbf{SASRec} & 0.1217 & 0.0627 & 0.1209 & 0.0555 & 0.1165 & 0.0494 \\
\textbf{BIGRec} & 0.0842 & 0.0334 & 0.0100 & 0.0044 & 0.0978 & 0.0416 \\
\midrule
\textbf{SFT} & 0.1895 & 0.0794 & 0.1847 & 0.0550 & 0.1282 & 0.0431 \\
\cellcolor{gray!16}\textbf{RosePO-h} & \cellcolor{gray!16}\textbf{0.2000} & \cellcolor{gray!16}\textbf{0.0815} & \cellcolor{gray!16}\textbf{ 0.1930} & \cellcolor{gray!16}\textbf{0.0575} & \cellcolor{gray!16}\textbf{0.1543} &\cellcolor{gray!16}\textbf{0.0516} \\
\bottomrule
\end{tabular}
}}
\label{tab:overall_all-ranking_performance}
\end{table}